\documentstyle[aps,psfig,eqsecnum]{revtex}
\begin{document}

\draft

\title{A Note On Holographic Ward Identities} 

\author{Steven Corley\thanks{scorley@phys.ualberta.ca}}

\address{Theoretical Physics Institute,
Department of Physics, University of Alberta,
Edmonton, Alberta, Canada T6G 2J1}

\maketitle
\begin{abstract}
In this note we show how Ward identities may be derived
for a quantum
field theory dual of a string theory using the $AdS$/CFT
correspondence.  In particular associated with any gauge
symmetry of the bulk supergravity theory there is a corresponding
constraint equation.  Writing this constraint in 
Hamilton-Jacobi form gives a generating functional for
Ward identities in the dual QFT.  We illustrate the
method by considering various examples. 
\end{abstract}
\pacs{}

\section{Introduction} 

The development of the $AdS$/CFT correspondence 
\cite{mald,gkp,witten}, see \cite{agmoo} for a review, has shown that
the ``radial'' coordinate of anti-de Sitter spacetime, by which
we mean a coordinate transverse to the timelike boundary, corresponds
in the dual CFT to an energy scale.  In particular the UV/IR
relations discussed in \cite{suss,peet} illustrate how large energies
in the CFT correspond to being near the boundary of AdS, and
conversely how small energies in the CFT correspond to being
far from the boundary of AdS.  This has led to the realization
that ``radial'' evolution in the bulk corresponds to renormalization
group flow in the boundary, i.e., that the bulk equations of
motion are the renormalization group equations of the boundary
correlators.  This idea has been elucidated in various cases
involving the duality between type IIB string theory on $AdS_5
\times S^5$ and 4-dimensional ${\cal N}=4$ SYM, 
\cite{akh,alv,fre,gir,gir2,por,ske,dew,dew2}.
There have
further been more general analyses given in \cite{bal,deb,li}.  In particular
de Boer, Verlinde, and Verlinde have given an explicit construction
of the renormalization group equations from the equations of motion
for a set of scalar fields coupled to gravity.  Their method consists
of first rewriting the equations of motion for the coupled system
in Hamilton-Jacobi form using a radial coordinate as the evolution
parameter.  Diffeomorphism invariance gives rise to the Hamiltonian
and vector constraints, the former of which \cite{deb} use
prominently to construct the dual QFT renormalization group equations.
Their analysis however makes no use of the vector constraints, so
it is natural to wonder what role it plays in the grand scheme of
things.  We show below that it, and more generally any constraint
arising from a gauge symmetry, gives rise to Ward identities in 
the dual QFT.  Conversely this means that in trying to reconstruct
the bulk spacetime from the boundary QFT (see, eg., \cite{deH,bal2,lif}), 
Ward identities in
the QFT will play an important role in constructing the bulk
constraint equations.

The subject of Ward identities in $AdS$/CFT has arisen as a
useful check on the correspondence, i.e., one may compute
CFT correlators from supergravity on anti-de Sitter spacetime
and check that they satisfy the required Ward identities.
Such checks have been carried out in various cases.  For
the duality between 4d, ${\cal N}=4$ super Yang-Mills (SYM)
theory and type IIB string theory on $AdS_5 \times S^5$
a check on the Ward identity of the $R$-current correlators
$\langle J^{a}_{\mu}(x) J^{b}_{\nu}(y) J^{c}_{\rho}(z) \rangle$
of the SYM theory has been carried out in \cite{Freedetal,CNSS}.
\cite{Freedetal} 
also showed that the dual QFT correlator
$\langle J^{a}_{\mu}(x) {\cal O}^{I}(y) {\cal O}^{J}(z) \rangle$
(where 
${\cal O}^{I}(x)$ is a gauge invariant composite scalar) 
satisfies the appropriate Ward identity, with similar
results for spinors replacing the above scalars reported
in \cite{muck}.  One also has Ward identites involving
the stress-tensor of the dual CFT and indeed \cite{liu}
showed that correlators of the form 
$\langle T_{\mu \nu}(x) {\cal O}(y) {\cal O}(z) \rangle$
computed in the supergravity approximation satisfy the
appropriate identity.
Our purpose in this note is to show that indeed these Ward
identities, as well as a more general set of identities,
follow quite simply from the supergravity constraint equations.
Specifically, corresponding to any gauge symmetry on the supergravity side
of the duality is a constraint equation.  Rewriting this constraint
equation in Hamilton-Jacobi form as a functional of the boundary
values of the supergravity fields, taking the appropriate functional
derivatives of the resulting equation with respect to the boundary
values of the fields, and evaluating on the appropriate background
yields precisely the dual QFT Ward identities.

\section{Ward identities}

To begin, let's review the prescription \cite{gkp,witten} 
for generating correlators
of operators in the dual quantum field theory from supergravity
on an asymptotic $AdS_{d+1}$ space.  Following \cite{deb} we work
with the Euclidean signature case and assume that the manifold
is topologically an $d+1$-ball as is Euclidean $AdS_{d+1}$, but 
that geometrically
the space need only be asymptotic to $AdS_{d+1}$.  We write the metric
in $ADM$ form, appropriate for radial evolution, as\footnote{We use
Greek letters $\mu, \nu, ...$ to denote $(d+1)$-dimensional coordinate
indices and Latin letters $i, j, ...$ to denote $d$-dimensional coordinate
indices.}
\begin{equation}
ds^2 = N^2 dr^2 + h_{ij}(x,r)(dx^i + N^i dr)(dx^j + N^j dr).
\label{metric}
\end{equation}
In particular we will eventually choose the ``radial'' gauge conditions
$N=1$ and $N^i=0$, and furthermore take 
$h_{ij}(r,x) = \delta_{ij} \exp(-2 \lambda(r))$ as in the ``kink''
solutions\footnote{Some of the scalars $\phi^I$ to be introduced shortly
will also be nonzero in the kink background and depend only
on $r$, i.e., $\phi^{I}(x,r) = \phi^{I}(r)$.} considered in 
\cite{akh,alv,fre,gir,gir2,por,ske,dew,dew2}, 
but for now leave the metric in the
general form (\ref{metric}).
The prescription for generating correlators associates to each field
$\phi_i$ in the string theory spectrum a corresponding local operator
${\cal O}^i$ in the QFT such that the relation (in
Euclidean signature)
\begin{equation}
Z[\phi_{i,0}] = 
\Bigl\langle e^{\int_{\partial} \phi_{i,0} {\cal O}^i}
\Bigr\rangle
\label{correspondence}
\end{equation}
holds,
where the left-hand-side is the string theory partition function with the
background fields $\phi_i$ turned on with boundary values
$\phi_{i,0}$.  In the large $N$, large 't Hooft coupling limit,
the partition function can be approximated 
by $\exp(-S_{eff}[\phi_{i,0}])$ where
$S_{eff}$ is the low energy supergravity effective action evaluated on 
the solutions to 
it's equations of motion subject to the boundary conditions
$\phi_i |_{\partial} = \phi_{i,0}$.  To regulate the action 
$S_{eff}[\phi_{i,0}]$ we take the boundary at finite $r$.
On the right-hand-side of (\ref{correspondence})
the expectation value of the given exponential is taken in the dual
quantum field theory, with $\phi_{i,0}$ acting as a source for
the QFT operator ${\cal O}^i$.  In particular the boundary value
of a bulk scalar, spinor,
massless vector, massless gravitino, and massless graviton acts 
respectively as the source for
a scalar, spinor, conserved current, conserved SUSY 
current, and stress tensor in the dual QFT.
To construct correlators for the QFT fields ${\cal O}^i(x)$ now,
one need just functionally differentiate the above relation
(\ref{correspondence}) with respect to the source $\phi_{i,0}$ the
appropriate number of times, eg.,
\begin{eqnarray}
\langle {\cal O}^{i_1}(x_1) \cdots {\cal O}^{i_n}(x_n) \rangle 
=  - \frac{\delta}{\delta \phi_{i_1,0}(x_1)} \cdots
\frac{\delta}{\delta \phi_{i_n,0}(x_n)} S_{eff}[\phi_{i,0}] |_{\phi_{i,0}=0}.
\end{eqnarray}

\subsection{Vector case:}
\label{Vectorcase}

According to the above dictionary between bulk and boundary fields,
to derive QFT Ward identities the
dual supergravity theory must contain some massless vectors, gravitinos,
and a graviton.  For simplicity we shall ignore the gravitino, commenting
on it later, and concentrate
on the other two fields.  In particular let's first consider a
real scalar field
minimally coupled 
to a non-Abelian
vector field with gauge group $SO(N)$, and take the scalars in
the representation $t^{IJ}_a$ where the matrices $t^{IJ}_a$ are 
imaginary and antisymmetric.  There are of course more relevant
systems to consider as far as the $AdS$/CFT duality is concerned,
eg., 5d gauged ${\cal N}=8$ supergravity, but our purpose here is
only to illustrate the method of generating Ward identities in 
the dual QFT from the supergravity description of the theory, and
with this goal in mind the simple system we consider here, and in
the next section, suffices.
The action for 
this system in a curved spacetime
is
\begin{eqnarray}
S_{\phi} + S_{A}  =   \int d^{d+1} x \sqrt{g} \bigl(
\tilde{\nabla}_{\mu} \phi^I
\tilde{\nabla}^{\mu} \phi^I 
+ V(\phi)
+ \frac{1}{4 g_{SG}^2} F^{a}_{\mu \nu}
F^{a \mu \nu} \bigr)
\end{eqnarray} 
where $V(\phi)$ is a potential for the scalars,
$F^{a}_{\mu \nu} = \nabla_{\mu} A^{a}_{\nu} - \nabla_{\nu} A^{a}_{\mu}
+c_{bc}^{\hspace{1em} a} 
A^{b}_{\mu} A^{c}_{\nu}$, $\tilde{\nabla}_{\mu} \phi^I =
\nabla_{\mu} \phi^I - i A^{a}_{\mu} t^{IJ}_a \phi^J$, 
and $[t_a, t_b] = 
c_{ab}^{\hspace{1em} c} t_c$.  
Including the gravitational part (see \ref{EHaction}) we assume
that the gravity-scalar subsystem possesses kink solutions of
the form mentioned after (\ref{metric}).

To construct the Hamiltonian appropriate for radial
evolution we decompose the the metric as $g_{\mu \nu} = h_{\mu \nu}
+ n_{\mu} n_{\nu}$ where $n_{\mu}$ is the unit normal to
an $n-1$ dimensional surface of constant $r$ and
$n^{\mu} h_{\mu \nu} = 0$.  Given a vector
$r^{\mu}$ satisfying $r^{\mu} \nabla_{\mu} r = 1$ then
$n_{\mu}$ may be decomposed as $n^{\mu} = \frac{1}{N}(r^{\mu}
- N^{\mu})$.
Decomposing the
vector potential as $A^{a}_{\mu} = n_{\mu} V^a/N + \tilde{A}^{a}_{\mu}$
where $n^{\mu} \tilde{A}^{a}_{\mu} = 0$ we find that
\begin{equation}
H = \int d^{d}x \sqrt{h} \bigl( N {\cal H} +  N^{\mu}
{\cal H}_{\mu} -  V^{a} G_a \bigr)
\label{Hamiltonian}
\end{equation}
where
\begin{eqnarray}
{\cal H} & = & {\cal H}_{A} + {\cal H}_{\phi} \\
{\cal H}_{A} & = &
\frac{g_{SG}^2}{2} \tilde{\pi}^{a \mu} \tilde{\pi}^{a}_{\mu}
- \frac{1}{4 g_{SG}^2} \tilde{F}^{a}_{\mu \nu} \tilde{F}^{a \mu \nu} \\
{\cal H}_{\phi} & = & \frac{1}{2} \Bigl(
\pi^{I} \pi^{I}
- \tilde{D}_{\mu} \phi^I \tilde{D}^{\mu} \phi^I
- V(\phi) \Bigr) \\
{\cal H}_{\nu} & = & \tilde{\pi}^{a \mu} D_{\nu} \tilde{A}^{a}_{\mu}
-D_{\mu} (\tilde{\pi}^{a \mu} \tilde{A}^{a}_{\nu}) 
+ \pi^I D_{\nu} \phi^I 
\label{svvector} \\
G_a & = & D_{\mu} \tilde{\pi}^{a \mu} - c_{ab}^{\hspace{1em} c}
\tilde{A}^{b}_{\mu} \tilde{\pi}^{c \mu} 
+  i t^{IJ}_a \phi^I \pi^J
\label{gauge} \\
\tilde{\pi}^{a \mu} & = & \frac{1}{\sqrt{h}} \frac{\delta S}
{\delta \dot{\tilde{A}}^{a}_{\mu}}, \hspace{1em}
\pi^{I} = \frac{1}{\sqrt{h}} \frac{\delta S}
{\delta \dot{\phi}^{I}}
\end{eqnarray}
and $\dot{\phi}^I := {\cal L}_r \phi^I$,
$\dot{\tilde{A}}_{\mu}^a := {\cal L}_r \tilde{A}_{\mu}^a$, and
$D_{\mu}$ is the $h_{\mu \nu}$ metric compatible connection.
We will bring in the Einstein-Hilbert term for gravity momentarily
but this will not affect the $V^{a} G_a$ part of the Hamiltonian,
therefore we have the constraint $G_a(x) = 0$.

To arrive at the QFT Ward identities for the conserved currents
$J^{ai}(x)$ and scalars ${\cal O}^I(x)$ dual to the supergravity
vector $\tilde{A}^{a}_{i}(r,x)$ (we are now working in the radial
gauge $N=1$ and $N^i = 0$ so that $\tilde{A}^{a}_r(r,x)=0$)
and scalar $\phi^I(r,x)$ respectively
now consists of two steps.
The first is to recall from classical mechanics the Hamilton-Jacobi
relations
\begin{equation}
\tilde{\pi}^{a \mu} = \frac{1}{\sqrt{h}} \frac{\delta S}
{\delta \tilde{A}^{a}_{\mu}}, \hspace{1em}
\pi^{I} = \frac{1}{\sqrt{h}} \frac{\delta S}
{\delta \phi^{I}}
\label{HJmomenta}
\end{equation}
where the action $S$ has been evaluated on a solution to the
equations of motion and is viewed as a functional of the
boundary values of the fields 
$\tilde{A}^{a}_{i}(r_b,x)$ and $\phi^I(r_b,x)$ (the boundary being any
surface of constant $r=r_b$).  Substituting into
the constraint arising from the non-Abelian gauge invariance
(\ref{gauge})
results in
\begin{equation}
\sqrt{h} D_{i} \Bigl( \frac{1}{\sqrt{h}} 
\frac{\delta S}{\delta \tilde{A}^{a}_i} \Bigr) = 
c_{ab}^{\hspace{1em} c}
\tilde{A}^{b}_{i} \frac{\delta S}
{\delta \tilde{A}^{c}_i} - i t^{IJ}_a  
\phi^I \frac{\delta S}{\delta \phi^J}. 
\label{gaugeHJ}
\end{equation}

The second step to derive the QFT Ward identities is to use
the correspondence (\ref{correspondence}).  In particular, 
identifying\footnote{The expectation values depend on the
boundary radius $r_b$ although we don't show this dependence
explicitly.}
\begin{equation}
\langle J^{ai}(x) \rangle_{\tilde{A},\phi} = 
- \frac{\delta S}{\delta \tilde{A}^{a}_i(r_b,x)}, \hspace{1em}
\langle {\cal O}^I(x) \rangle_{\tilde{A},\phi} = 
- \frac{\delta S}{\delta \phi^I(r_b,x)}
\end{equation}
then we find the generating functional for Ward identities
\begin{equation}
\sqrt{h} D_{i} \Bigl( \frac{1}{\sqrt{h}} 
\langle J^{ai}(x) \rangle_{\tilde{A},\phi} 
\Bigr) = 
c_{ab}^{\hspace{1em} c}
\tilde{A}^{b}_{i}(r_b,x)
\langle J^{ci}(x) \rangle_{\tilde{A},\phi} 
- i t^{IJ}_a  
\phi^I(r_b,x) 
\langle {\cal O}^J(x) \rangle_{\tilde{A},\phi}
\label{Agen}
\end{equation}
where the subscript $\tilde{A},\phi$ denotes that the expectation
values are still being taken in the presence of the background sources.
Evaluating on a kink solution\footnote{Note that this requires
that $t^{IJ}_a (\phi^I \partial_r \phi^J - \phi^J \partial_r \phi^I)=0$.}
\begin{equation}
h_{ij}(x,r) = \delta_{ij} e^{-2 \lambda(r)}, \hspace{1em} \phi^I(x,r)
= \phi^I (r), \hspace{1em} \tilde{A}^{a}_{i}=0
\label{kink}
\end{equation}
we find that in general not all currents will be conserved, i.e.,
\begin{equation}
\partial_{i}   
\langle J^{ai}(x) \rangle_{\phi} =
- i t^{IJ}_a  
\phi^I(r_b) 
\langle {\cal O}^J(x) \rangle_{\phi},
\end{equation}
where the subscript $\phi$ denotes that only some of the scalar sources
are nonvanishing in the expectation values.  From the supergravity
perspective this nonconservation of the currect arises from 
spontaneous symmetry breaking by the background solution.  From
the dual QFT perspective however it arises because the theory has
been deformed away from it's CFT critical point, explicitly breaking
the global conservation law to a subgroup.

It is straightforward now to generate Ward identities for
$n$-point correlators by functionally differentiating (\ref{Agen})
with respect to the boundary values $\tilde{A}^{a}_i(r_b,x)$ and
$\phi^{I}(r_b,x)$ (and of course $h_{ij}(r_b,x)$ as well).
For example, differentiating (\ref{Agen}) $n$ times with respect
to $\phi^{I}$ and evaluating on a kink solution (\ref{kink})
yields
\begin{eqnarray}
\partial_{i}  
\langle J^{ai}(x) {\cal O}^{I_1}(x_1) \cdots {\cal O}^{I_n}(x_n) 
\rangle_{\phi} 
& = & 
- i \sum_{i=1}^{n}  t^{I_{i} J}_a \delta^{(d)}(x-x_i) 
\langle {\cal O}^J(x) {\cal O}^{I_1}(x_1) \cdots
\widehat{\cal O}^{I_i}(x_i) \cdots {\cal O}^{I_n}(x_n) \rangle_{\phi}
\nonumber \\
& - & i t^{IJ}_a  
\phi^I(r_b) 
\langle {\cal O}^J(x) {\cal O}^{I_1}(x_1) \cdots
{\cal O}^{I_n}(x_n) \rangle_{\phi}
\label{currentscalar}
\end{eqnarray}
where the notation $\widehat{\cal O}^I$ denotes that the 
operator is missing from the
expectation value.  For a pure $AdS$ background, i.e., no scalars
turned on, this reduces to the standard Ward identity which
was checked in \cite{Freedetal} to hold for $n=2$ in
the $\epsilon << 1$ limit where 
$r_b = -\ln \epsilon$.  It was noted in
\cite{Freedetal} that the prescription originally given in
\cite{witten} for computing 2-point functions did not yield
results consistent with the Ward identities.  \cite{Freedetal}
suggested another method for computing 2-point functions
along the lines of \cite{gkp} which involved solving the
equations of motion subject to boundary conditions placed
at a boundary $r=- \ln \epsilon$ for $\epsilon << 1$.  In the
Hamilton-Jacobi formalism used here this procedure comes out
naturally.

In a similar way one may derive the Ward identity for the
currents
\begin{eqnarray}
\partial_{i}   
\langle J^{ai}(x) J^{a_1 i_1}(x_1) \cdots J^{a_n i_n}(x_n) 
\rangle_{\phi} 
& = & 
\sum_{l=1}^{n}  c^{a a_l c} \delta^{(d)}(x-x_l) 
\langle J^{ci}(x) J^{a_1 i_1}(x_1) \cdots
\widehat{J}^{a_l i_l} (x_l) \cdots J^{a_n i_n}(x_n) \rangle_{\phi}
\nonumber \\
& - & i t^{IJ}_a  
\phi^I(r_b) 
\langle {\cal O}^J(x) J^{a_1 i_1}(x_1) \cdots
J^{a_n i_n}(x_n) \rangle_{\phi}
\label{npointcurrent}
\end{eqnarray}
where we have evaluated on the kink background (\ref{kink}).
For the case of $AdS$ background with no scalars turned on this
Ward identity was checked in \cite{Freedetal,CNSS} for $n=2$
and $r_b = 1/\epsilon$.

Another case of interest is to 
add a Chern-Simons term for the vector field $A^{a}_{\mu}$.
For the particular case of type IIB supergravity on
$AdS_5 \times S^5$ such an interaction term is produced
and gives rise, as first pointed out by Witten, to the 
three point current correlator anomaly
in the dual ${\cal N}=4$ SYM theory.  Witten further showed
how to compute the anomaly in a very simple manner.  Here
we consider the Chern-Simons term from the Hamilton-Jacobi
point of view and show that the anomaly again follows quite
easily.  Specifically consider the 5d Chern-Simons action
\begin{equation}
S_{CS} = \alpha \int d^{5}x d_{abc} 
\epsilon^{\mu \nu \lambda \rho \sigma}
A_{\mu}^a \partial_{\nu} A_{\lambda}^b \partial_{\rho} A_{\sigma}^c
\end{equation}
where $d_{abc} = Tr(\{t_a, t_b\} t_c)$. 
Repeating the above analysis
with the action $S_{\phi} +S_{A} + S_{CS}$ gives rise to the Chern-Simons 
contribution to the gauge constraint
\begin{equation}
G_{a,CS} = 
- \alpha 
\frac{1}{\sqrt{h}} \epsilon^{ijkl}(d_{abc} \partial_{i}
\tilde{A}_{j}^b \partial_{k} \tilde{A}_{l}^c
+ 2 d_{bcd} c_{ea}^{\hspace{1em} d} 
\tilde{A}_{i}^b \partial_{j} \tilde{A}_{k}^c
\tilde{A}_{l}^e )
\end{equation}
where the 4-dimensional epsilon symbol satisfies
$\epsilon^{\mu \nu \lambda \rho \sigma} = 5 N \epsilon^{[\mu \nu \lambda
\rho} n^{\sigma]}$.
Making the substitution (\ref{HJmomenta}) for the conjugate momenta
and functionally differentiating the above constraint two more
times with respect to $\tilde{A}^{a}_{i}$ gives rise to the Ward
identity (\ref{npointcurrent}) for $n=2$ plus the anomaly term
\begin{equation}
\frac{\partial}{\partial x^i} \langle J^{ai}(x) J^{bj}(y) 
J^{ck}(z) \rangle |_{anomaly} 
= - 2 \alpha d^{abc} \epsilon^{ijkl} 
\frac{\partial}{\partial x^{i}} \delta^{(d)}(x-y)
\frac{\partial}{\partial x^{l}} \delta^{(d)}(x-z).
\end{equation}
For a pure $AdS$ background this agrees with the anomaly calculation 
in the dual
${\cal N}=4$ SYM theory provided $\alpha = i(N^2 -1)/(96 \pi^2)$.

Including spinors and gravitinos is slightly more subtle than
the cases discussed so far as the action principles are first 
order in derivatives.  A consequence of this, see
\cite{sfe,muck},
is that one cannot fix all
components of the spinor at the boundary {\it and} demand 
regularity of the solution in the interior, but rather 
one must sacrifice half
of the spinor components at the boundary in order to keep
regularity of the solution.  To put it differently, half
of the spinor components must be thought of as the fields,
or ``coordinates'',
and the other half as the conjugate momenta.  Which components
to choose as the field variables is not arbitrary though.
For example, the solution to the free massive spinor equations
of motion on $AdS$ gives rise to the relation \cite{muck}
\begin{equation}
\psi^{+}(\epsilon,k) = -i \frac{k_i \gamma_i}{k} 
\frac{K_{m-1/2}(k \epsilon)}
{K_{m+1/2}(k \epsilon)} \psi^{-}(\epsilon,k)
\end{equation}
where $\gamma_i$ are the usual Dirac gamma matrices and the spinor
components $\psi^{\pm}$ are defined by 
$\psi^{\pm} = (1/2)(1\pm \gamma_0) \psi$.  For $m>0$ we see that
$\psi^{+}(\epsilon,k)$ goes to zero as $\epsilon \rightarrow 0$
for fixed $\psi^{-}(\epsilon,k)$.  Fixing $\psi^{+}(\epsilon,k)$
instead would result in a divergent $\psi^{-}(\epsilon,k)$.
Hence we conclude that $\psi^{-}$ should be viewed as the
field and $\bar{\psi}^{+}$ as the conjugate momentum,
where we have defined $\bar{\psi}^{\pm} = (1/2) \bar{\psi} (1 \pm \gamma_0)$,
and
similarly $\bar{\psi}^{-}$ as the field and $\psi^{+}$ as
the conjugate momentum.  For $m<0$ the opposite identifications
are made.
Another argument in favor of this identification was
given in \cite{arufrol} where it was shown for
$m>0$ that $\psi^{-}(\epsilon,x)$ 
transforms under a local representation of the conformal group
while $\psi^{+}(\epsilon,x)$ under a nonlocal representation.
It follows that only $\psi^{-}(\epsilon,x)$ can act as a source
for a quasi-primary operator in the dual CFT.  Using this identification
of fields and conjugate momenta it is straighforward to generalize
the derivation of the Ward identities above to include spinor
and gravitino fields.

\subsection{Removing the cutoff}

The derivation of the Ward identities presented so far has been in the
presence of a finite cutoff.  For massive
scalars ($m_{I}^2 > 0$) this
must be so as the space would not remain asymptotic to
$AdS_{d+1}$ in the $r \rightarrow \infty$ limit\footnote{Recall
that a scalar field with mass $m_{I}$ in an $AdS_{d+1}$ background
behaves near the $AdS_{d+1}$ boundary as $x_{0}^{d-\Delta}$
where
$\Delta = (1/2)(d + \sqrt{d^2 + 2 m^{2}_I})$ and $r=-\ln x_0$.}.
However, for tachyonic scalars,
ie., those with masses satisfying $-d^2/4 < m_{I}^2 < 0$,
we can remove the cutoff and rewrite the above Ward identities
in terms of renormalized fields.  One method that has been
used \cite{KleWit,deH} goes as follows.  Evaluate the action
at $r=-\ln \epsilon$ for $\epsilon << 1$.  The scalars vanish
as $\epsilon^{d - \Delta}$ (with 
$\Delta$ given in the previous footnote) in the $\epsilon 
\rightarrow 0$ limit,
the vector goes to a constant in this limit, and the metric
$h_{ij}$ diverges as $\epsilon^{-2}$, consequently the action
diverges.  Define the new fields
\begin{equation}
\tilde{\phi}^{I}(x_0,x) := (x_0)^{\Delta_I - d} \phi^{I}(x_0,x), \hspace{1em}
\tilde{h}_{ij}(x_0,x) := (x_0)^{2} h_{ij}(x_0,x)
\label{partren}
\end{equation}
where $r = -\ln x_0$.  More generally one can replace the
multiplicative factor of $x_0$ in (\ref{partren}) by
$f(x_0)$ so long as $f(x_0) \rightarrow x_0$ as $x_0 \rightarrow 0$.
The point being that the new fields $\tilde{\phi}^I$ and $\tilde{h}_{ij}$
approach constants as $x_0 \rightarrow 0$.
Expressing the action in terms of the boundary values of the
tilded fields allows one to isolate the divergences and remove
them by adding counterterms.  Furthermore the generating functional
of Ward identities (\ref{Agen})
when rewritten in terms of tilded fields takes
the same form as in terms of the original fields, but now everything
is finite and the $\epsilon \rightarrow 0$ limit can be taken.

Following \cite{deb} we may further rewrite the Ward identities
as functions of the background kink solution evaluated at a
finite radius (as opposed to leaving them as functions of the
asymptotic values of the tilded fields defined above).  This simply 
corresponds to choosing a different renormalization scale in
the dual QFT.
Define therefore the renormalized fields
\begin{equation}
\phi^{I}_{R} := \phi^I(x_0), \hspace{1em} h_{ij,R} := h_{ij}(x_0)
\end{equation}
for some $x_0$.  The fields $\phi^{I}_R$ and $h_{ij,R}$
can be expressed as functions of the asymptotic values of the
fields $\tilde{\phi}^{I}(x_0)$ and $\tilde{h}_{ij}(x_0)$, and
vice versa.
In particular we have $\phi^{I}_0 = \phi^{I}_0 (\tilde{\phi}^{J}_R)$
where $\tilde{\phi}^{J}_0$ denotes the asymptotic value of
$\tilde{\phi}^{I}(x_0)$.  Under an infinitesimal global gauge
transformation this relation yields
\begin{equation}
\epsilon^a t^{IJ}_a \phi^{J}_0 = \phi^{I}_0 (\epsilon^a t^{JK} 
\tilde{\phi}^{K}_R) \approx  \epsilon^a \frac{\partial \phi^{I}_0}
{\partial \phi^{J}_R} t^{JK}_a \phi^{K}_R.
\label{gaugex}
\end{equation}
Defining the renormalized operators ${\cal O}^{I}_R$ by
\begin{equation}
{\cal O}^{I}_R := \frac{\partial \tilde{\phi}^{J}_0}{\partial \phi^{I}_R}
{\cal O}^{J}
\end{equation}
and using the relation (\ref{gaugex}) it follows that the Ward
identity (\ref{currentscalar}) (by which we really mean 
(\ref{currentscalar}) expressed in terms of the tilded fields), and more 
generally any Ward identity involving
the scalars ${\cal O}^I$ and currents $J^{ai}$, takes the same
form, i.e.,
\begin{eqnarray}
\partial_{i}  
\langle J^{ai}(x) {\cal O}^{I_1}_R(x_1) \cdots {\cal O}^{I_n}_R(x_n) 
\rangle_{\phi} 
& = & 
- i \sum_{i=1}^{n}  t^{I_{i} J}_a \delta^{(d)}(x-x_i) 
\langle {\cal O}^{J}_R(x) {\cal O}^{I_1}_R(x_1) \cdots
\widehat{\cal O}^{I_i}_R(x_i) \cdots {\cal O}^{I_n}_R(x_n) \rangle_{\phi}
\nonumber \\
& - & i t^{IJ}_a  
\phi^{I}_R 
\langle {\cal O}^{J}_R(x) {\cal O}^{I_1}_R(x_1) \cdots
{\cal O}^{I_n}_R(x_n) \rangle_{\phi}
\end{eqnarray}
where the expectation values are now expressed as functions of
$\phi^{I}_R$.

\subsection{Inclusion of the metric:}

We can further derive QFT Ward identities associated 
with translation invariance by looking at the constraint
arising from varying the shift vector $N^{\mu}$.  We must
first however include the gravitational part of the action
given by the Einstein-Hilbert Lagrangian
\begin{equation}
S_{EH} = \frac{1}{2 \kappa^2} \int d^{d+1} x \sqrt{g} R.
\label{EHaction}
\end{equation}
The Hamiltonian for this system again takes the same form as
in (\ref{Hamiltonian}) with vanishing contribution to 
the constraint $G_a$ while
\begin{eqnarray}
{\cal H}_g & = & - \frac{1}{2 \kappa^2} (\pi^{\mu \nu} \pi_{\mu \nu} - 
\frac{1}{n-2}
\pi^2 + \tilde{R}) \\
({\cal H}_g)_{\nu} & = & - \frac{1}{\kappa^2} D_{\mu} 
\pi^{\mu}_{\hspace{0.5em} \nu} \label{gravvector} \\
\frac{1}{2 \kappa^2} \pi^{\mu \nu} & = & \frac{1}{\sqrt{h}} 
\frac{\delta S_{EH}}{\delta \dot{h}_{\mu \nu}}
\end{eqnarray}
where $\tilde{R}$ is the Ricci scalar associated with $h_{\mu \nu}$.
Combining the vector constraint for the Einstein-Hilbert action
(\ref{gravvector}) with that for the scalar-vector system
(\ref{svvector}) and replacing momenta as in (\ref{HJmomenta}) for
the scalars and vectors and 
\begin{equation}
\frac{1}{2 \kappa^2} \pi^{ij} = \frac{1}{\sqrt{h}} 
\frac{\delta S}{\delta h_{ij}}
\end{equation}
for the metric (evaluating in the radial gauge) results in 
\begin{equation}
-2 \sqrt{h} D_{j} \Bigl(\frac{1}{\sqrt{h}} 
\frac{\delta S}{\delta h^{i}_{\hspace{0.5em} j}} \Bigr)
+ \partial_{i} \phi^{I} \frac{\delta S}{\delta \phi^I}
+D_{i} \tilde{A}^{a}_{j} \frac{\delta S}{\delta \tilde{A}^{a j}}
- \sqrt{h} D_{j} \Bigl( \frac{1}{\sqrt{h}} 
\frac{\delta S}{\delta \tilde{A}^{a}_{j}} \tilde{A}^{a}_{i} \Bigr) = 0.
\label{vectorWard}
\end{equation}
Assuming that the dual QFT stress tensor couples to the boundary
value of the metric as
$\frac{1}{2} \int d^d x T_{ij} h^{ij}$, then evaluating
(\ref{vectorWard}) on a kink background implies that the
stress tensor is conserved, i.e., 
$\partial_i \langle T^{ij}(x) \rangle_{\phi}
=0$.
To derive the Ward identities we need only functionally differentiate
as before, eg.,
differentiating (\ref{vectorWard}) $n$ times with respect to
$\phi^{I_i}(x_i)$ and evaluating on the kink background
results in 
\begin{equation}
\frac{\partial}{\partial x^{j}} \langle T^{j}_{\hspace{0.5em} i}(x)
{\cal O}^{I_1} (x_1) \cdots {\cal O}^{I_n}(x_n) \rangle =
\sum_{k=1}^{n} \frac{\partial}{\partial x^{i}} \delta^{(d)}(x-x_k)
\langle {\cal O}^{I_1} (x_1) \cdots {\cal O}^{I_k}(x)
\cdots  {\cal O}^{I_n}(x_n) \rangle.
\end{equation}
Differentiating instead with respect to the vector one finds instead
\begin{eqnarray}
\frac{\partial}{\partial x^{j}} \langle T^{i}_{\hspace{0.5em} j}(x)
J^{a_1 i_1}(x_1) \cdots J^{a_n i_n}(x_n) \rangle & = & \sum_{k=1}^{n}
\Bigl( \frac{\partial}{\partial x^{i}} \delta^{(d)}(x-x_k)
\langle J^{a_1 i_1}(x_1) \cdots J^{a_k i_k}(x)
\cdots J^{a_n i_n}(x_n) \rangle \nonumber \\ & - &
\frac{\partial}{\partial x^{j}} \bigl(\delta^{(d)}(x-x_k) \delta^{i_k}_i 
\langle J^{a_1 i_1}(x_1) \cdots J^{a_k j}(x_k)
\cdots J^{a_n i_n}(x_n) \rangle \bigr) \Bigr). 
\end{eqnarray}

To derive the Ward identities for the renormalized fields one can
follow the procedure outlined previously.

\section{Conclusions}

We have shown that Ward identities in the quantum field theory dual
of a string theory on an asymptotically $AdS$ space follow quite
naturally from the supergravity constraint equations.  Specifically
associated to any gauge symmetry of the bulk supergravity 
theory there is a corresponding constraint equation.  The generating
functional of QFT Ward identities follows after rewriting this
constraint equation in Hamilton-Jacobi form.  We have considered
in particular the examples of a non-Abelian gauge invariance and
diffeomorphism invariance in the bulk and have computed the corresponding
Ward identity generating functionals.

It is not surprising that there is a corresponding
Lagrangian approach to deriving
these Ward identities as well.  Namely varying the $AdS$/CFT relation
(\ref{correspondence}) under a supergravity gauge transformation leaves the
supergravity side of the relation invariant (up to possible anomalies,
one case of which was discussed in section \ref{Vectorcase}).  The
dual QFT expectation value however is not invariant and gives
rise to a generating functional for the corresponding Ward identities.
This argument was already implicit in Witten's computation of the 
$R$-current anomaly for 4d, ${\cal N}=4$ super Yang-Mills theory and
more recently was used explicitly in \cite{deH} to derive the
generating functional (\ref{vectorWard}) (without the vector field).
One advantage in the approach used here to derive QFT Ward identities
is that it sheds some light on the inverse problem of reconstructing
the bulk spacetime from the dual QFT.  Specifically in the approach
of this paper one can see more explicitly how the QFT Ward identities
correspond to the bulk supergravity constraint equations.  One final
comment is that it would be interesing to find a Lagrangian approach 
to deriving
the QFT renormalization group equations dual to the Hamilton-Jacobi
approach used in \cite{deb}, which one would naturally expect must
exist especially in light of the dual approaches to deriving Ward
identities.

\section*{Acknowledgements}
We would like to thank E. Verlinde for some helpful email correspondence.
This work was supported by the Natural Sciences and Engineering
Research Council of Canada at the University of Alberta.

\end{document}